\documentclass[showpacs,preprintnumbers,amsmath,amssymb,twocolumn,superscriptaddress,prb]{revtex4}
\usepackage{times}
\usepackage{graphicx}
\usepackage{amsfonts}
\usepackage{amsmath, amsthm, amssymb}
\usepackage{dsfont}
\usepackage{subfigure}

\newcommand{\etal}{\emph{et al. }}

\begin{document}
\title{Triplet proximity effect and odd-frequency pairing in graphene}
\author{Jacob Linder}
\affiliation{Department of Physics, Norwegian University of
Science and Technology, N-7491 Trondheim, Norway}
\author{Annica M. Black-Schaffer}
\affiliation{NORDITA, Roslagstullsbacken 23, SE-106 91 Stockholm, Sweden}
\author{Asle Sudb{\o}}
\affiliation{Department of Physics, Norwegian University of
Science and Technology, N-7491 Trondheim, Norway}

\date{Received \today}
\begin{abstract}
We study the interplay between proximity-induced superconductivity and ferromagnetism in graphene by self-consistently solving the 
Bogoliubov-de Gennes equations on the honeycomb lattice. We find that a strong triplet proximity effect is generated in graphene, 
leading to odd-frequency pairing correlations. These odd-frequency correlations are clearly manifested in the local density of states 
of the graphene sheet, which can be probed via STM-measurements. Motivated by recent experiments on S$\mid$N$\mid$S graphene Josephson 
junctions, we also study the spectrum of Andreev-bound states formed in the normal region due to the proximity effect. Our results may 
be useful for interpreting spectroscopic data and can also serve as a guideline for future experiments.
\end{abstract}

\maketitle

%
\textit{Introduction}. Graphene \cite{novoselov_science_04} constitutes a new exciting setting for studying the interplay 
between 
different 
types of long-range orders, such as ferromagnetism and superconductivity. Although the intrinsic appearance of both of these 
phenomena only 
occurs under special circumstances in graphene (see e.g.~Refs.~\onlinecite{Ohldag07, Cervenka09, Esquinazi08}), they can 
always be induced 
via proximity to host materials with the desired properties. 
The study of how the peculiar electronic properties \cite{castroneto_rmp_09} of graphene interact with superconducting 
correlations has 
recently attracted much attention both theoretically \cite{beenakker_prl_06, titov_prb_06, beenakker_rmp_08} and
 experimentally.\cite{heersche_nature_07, du_prb_08, ojeda_prb_09}
Such studies are done by depositing two superconducting leads on graphene in order to create a graphene S$\mid$N$\mid$S 
Josephson junction. By also exposing the N region to a ferromagnetic host, a hybrid  S$\mid$F$\mid$S junction is constructed, 
which then will offer an excellent platform in which to study the interplay between ferromagnetism and superconductivity. 
Previous work on such hybrid structures have reported on interesting effects by studying its transport properties via a 
scattering matrix approach.\cite{linder_prl_08, moghaddam_prb_08} 
However, this formalism does not include the full extent of the superconducting proximity effect, as it does not self-consistently 
solve for the superconducting order parameter inside the junction. A self-consistent solution, on the other hand, will explicitly 
include the Cooper pair depletion in S, and the corresponding leakage into N, near the interfaces. In addition, the scattering 
matrix approach cannot deduce the symmetry of the induced correlations in the non-superconducting region, nor calculate their 
manifestation in the local density of states (DOS). 
In particular, it remains be clarified how odd-frequency correlations \cite{berezinskii_jetp_74, balatsky} adapt to the unusual 
electronic environment of graphene and what their signature is in experimentally accessible quantities. The issue of odd-frequency 
pairing has recently generated much activity in the field of conventional F$\mid$S junctions,\cite{bergeret_rmp_05} but has not 
yet been addressed in the context of graphene.

In this paper, we present a self-consistent lattice-study of the interplay between ferromagnetism and superconductivity in graphene, 
with focus on the behavior of the local DOS to probe this interaction. Specifically, we investigate an S$\mid$F$\mid$S junction and 
the concomitant manifestation of odd-frequency pairing. To make contact with the current experimental status, we also consider an 
S$\mid$N$\mid$S junction and identify the appearance of subgap Andreev-bound states in the DOS. Our results should provide a 
guideline for spectroscopic measurements in graphene when superconductivity and/or ferromagnetism are induced by means of the 
proximity effect.
%
\begin{figure*}
\centering
\resizebox{0.90\textwidth}{!}{
\includegraphics{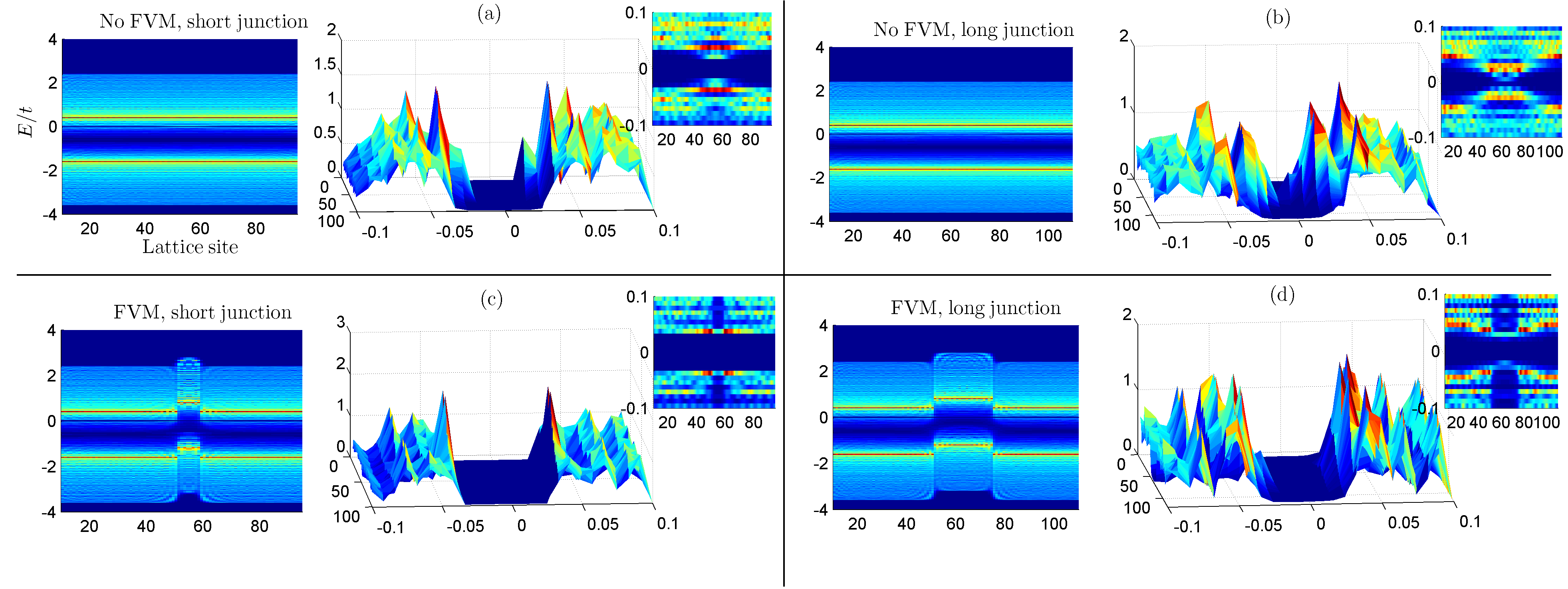}}
\caption{(Color online). Local DOS on a large energy scale (left panel) and near the gapped region (right panels) in an S$\mid$N$\mid$S 
graphene junction for a variety of lengths and FVMs. Large values of the DOS are indicated by a bright color, while small values are 
given by a dark color.
We have fixed $L_S=45$ sites and $\mu_S/t=0.6$, as well as set the superconducting phase difference to zero, $\phi=0$. 
(a): $\mu_N=\mu_S$ (no FVM), and $L_N = 8$ sites ($\xi \sim 25$ sites). (b): $\mu_N=\mu_S$ and $L_N = 24$ sites. 
(c): $\mu_N/t=0.2$ and $L_N = 8$ sites. (d): $\mu_N/t=0.2$ and $L_N = 24$ sites. As seen, in-gap bound states, below 
the gap edge, are formed and persist even in the long-junction regime.}
\label{fig:SNS_total}
\end{figure*}
%

%
\textit{Theory}. We start out with the following tight-binding lattice Hamiltonian for graphene
%
\begin{align}\label{eq:H}
\mathcal{H} &= -t\sum_{\langle\mathbf{i},\mathbf{j}\rangle,\sigma} (f_{\mathbf{i}\sigma}^\dag g_{\mathbf{j}\sigma} + g_{\mathbf{i}\sigma}^\dag f_{\mathbf{j}\sigma}) -\sum_{\mathbf{i}\sigma} \mu_{\mathbf{i}\sigma}(f_{\mathbf{i}\sigma}^\dag f_{\mathbf{i}\sigma} + g_{\mathbf{i}\sigma}^\dag g_{\mathbf{i}\sigma})\notag\\
&- \sum_\mathbf{i} U_\mathbf{i} (f_{\mathbf{i}\uparrow}^\dag f_{\mathbf{i}^\uparrow} f_{\mathbf{i}\downarrow}^\dag f_{\mathbf{i}\downarrow} + g_{\mathbf{i}\uparrow}^\dag g_{\mathbf{i}^\uparrow} g_{\mathbf{i}\downarrow}^\dag g_{\mathbf{i}\downarrow}),
\end{align}
and largely employ the notation and methods of Ref.~\onlinecite{blackschaffer}. Here 
$f_{\mathbf{i}\sigma}^\dagger$ ($g_{\mathbf{i}\sigma}^\dagger$) is the creation operator on the $A$ ($B$) site of the 
honeycomb lattice, $t \sim 2.5$~eV is the nearest neighbor hopping parameter, $\langle\mathbf{i},\mathbf{j}\rangle$ 
denotes summation over nearest neighbors, and $\sigma$ is the spin index.
Moreover, $\mu_{\mathbf{i}\sigma} = \mu_\mathbf{i} + \sigma h_\mathbf{i}$ is the spin-dependent chemical potential. We 
will assume that the native chemical potential $\mu_\mathbf{i}$ is a constant within each region of the junction (S or N/F). 
Experimentally, an overall chemical potential can be set in the whole sample by applying a back gate voltage. In addition, 
it is expected that some charge transfer takes place between graphene and the superconducting leads and, therefore, $\mu_S$ 
can sometimes be higher than $\mu_{N/F}$. $h_\mathbf{i}$ is the site-dependent exchange field, which regulates the 
ferromagnetic order induced by proximity to the ferromagnetic host material. Thus, $h_\mathbf{i}$ is only non-zero 
in the region between the two superconducting leads.
The last term in Eq.~(\ref{eq:H}) models the influence of the superconducting leads on graphene. The attractive on-site 
interaction $U_\mathbf{i}$ gives rise to $s$-wave superconductivity and this parameter is only non-zero in the S regions 
of the junctions. 

We perform a mean-field approximation on Eq.~(\ref{eq:H}), with the superconducting order parameter defined by
%
\begin{align}\label{eq:OPself}
\Delta_{\bf i} &=-U_{\bf i} [\langle f_{{\bf i}\downarrow}f_{{\bf i}\uparrow} \rangle + \langle g_{{\bf i}\downarrow}g_{{\bf i}\uparrow} \rangle]/2.
\end{align}
We have ignored spatial variations within one unit cell and used $\Delta_f=\Delta_g$. We further consider a geometry 
with translational invariance along the interfaces, i.e.~orthogonal to the direction of the junction. For concreteness, 
we focus on a zig-zag interface and Fourier-transform the eigenvectors in the $y$-direction. Please note that for an 
$s$-wave symmetry, the specific direction of the interface will not matter. After diagonalizing the problem, we arrive 
at the tight-binding Bogoliubov-de Gennes (BdG) equations for graphene
%
\begin{align}\label{eq:diag}
\sum_m &\begin{pmatrix}
\hat{H}_\uparrow(n,m) & \hat{\Delta}(n,m) \\
\hat{\Delta}^\dag(n,m) & -\hat{H}_\downarrow(n,m) \\
\end{pmatrix} \psi^\nu_n = E^\nu(k_y)\psi^\nu_n.
\end{align}
Here $n$ is the lattice site index along the junction and $k_y =2\pi l/(N_y a)$, where $l$ is an integer such that 
$k_y \in ]-\pi/a,\pi/a]$, and $a$ is the lattice constant. Moreover, 
$\psi^\nu_n = [u_n^\nu(k_y), y_n^\nu(k_y), v_n^\nu(k_y), z_n^\nu(k_y)]^\text{T}$ is generated by two copies (one 
for each sublattice) of the standard canonical Bogoliubov transformation. We have also introduced the following 
matrices
%
\begin{widetext}
\begin{align}
\hat{H}_\sigma(n,m) = \begin{pmatrix}
-(\mu_n+\sigma h_n)\delta_{nm} & -t[\delta_{nm} + 2\delta_{n+1,m}\cos(k_ya/2)] \\
-t[\delta_{nm} + 2\delta_{n-1,m}\cos(k_ya/2)] & -(\mu_n+\sigma h_n)\delta_{nm}\\
\end{pmatrix},\; \hat{\Delta}(n,p) = \begin{pmatrix}
\Delta_n\delta_{nm} & 0 \\
0 & \Delta_n\delta_{nm} \\
\end{pmatrix}.
\end{align}
\end{widetext}
For a self-consistent solution of the above equations, we first guess an initial $\Delta_n$, then find the corresponding 
eigenvalues and eigenvectors to Eq.~(\ref{eq:diag}), followed by a recalculation of $\Delta_n$ using the self-consistency 
criteria in Eq.~(\ref{eq:OPself}). This process is iterated until $\Delta_n$ no longer changes between subsequent 
iterations. We note in passing that for the case of a zero exchange field, Eq.~(\ref{eq:diag}) is particle-hole symmetric 
and it is then enough to solve for only the negative eigenvalues. However, this is not the case for a non-zero $h$, and 
we are therefore forced to calculate all the eigenvalues.

We are particularly interested in investigating the appearance of spin-triplet correlations in the system when the exchange 
field $h_i$ is non-zero. These triplet correlations are necessarily of an unusual nature, since the spatial symmetry of the 
superconducting order parameter is isotropic ($s$-wave). According to the Pauli-principle, it is possible to have superconducting 
correlations which are both isotropic and spin-triplet simultaneously as long as these have an \textit{odd-frequency symmetry}. 
The frequency dependence is obtained by Fourier-transforming the relative time-coordinate $(\tau-\tau')$, and it thus follows 
that such an odd triplet amplitude must be antisymmetric with respect to $(\tau-\tau')$. In effect, this amounts to a 
strong retardation effect since the correlator vanishes at equal times. The study of proximity-induced odd-frequency 
pairing has recently generated much interest in conventional metallic S$\mid$F systems, but has not yet been explored 
in graphene. To investigate the presence of odd-frequency pairing, we introduce \cite{halterman}
%
\begin{align}\label{eq:triplet}
\mathcal{F}_{f,{\bf i}}^t = \langle f_{{\bf i}\uparrow}(\tau)f_{{\bf i}\downarrow}(0) +  f_{{\bf i}\downarrow}(\tau)f_{{\bf i}\uparrow}(0)\rangle. 
\end{align}
As with the order parameter in Eq. (\ref{eq:OPself}), we define the effective odd-frequency correlator $\mathcal{F}_{{\bf i}}^t$ as the average between the expectation values on the two sublattices described by the $f$- and $g$-fermion operators. We will also be concerned with the local DOS $\mathcal{N}_{\bf i}(E)$ which in the low-temperature limit is obtained via the charge density $\rho$ as follows:
\begin{align}
\rho_{\bf i} = \int^0_{-\infty} \mathcal{N}_{\bf i}(E)\text{d}E = \sum_\sigma \langle f_{\mathbf{i}\sigma}^\dag f_{\mathbf{i}\sigma} + g_{\mathbf{i}\sigma}^\dag g_{\mathbf{i}\sigma}\rangle.
\end{align}

%
%
\textit{Results and Discussion.} In what follows, we present a numerical and self-consistent solution of the above equations. Let us 
first consider the S$\mid$N$\mid$S case, shown in Fig.~\ref{fig:SNS_total}. To model experimentally relevant scenarios, we consider 
both short and long junctions in addition to the presence or absence of a Fermi-vector mismatch (FVM) at the interface.
It may be instructive to start with a reminder of the analytical result for the Andreev bound-state energy inside a junction, 
obtainable in the ultrashort-junction regime where $L\ll\xi$, with $\xi$ being the superconducting coherence length. This relation 
reads $E = \Delta_0[1-D\sin^2(\phi/2)]^{1/2}$, where $D$ denotes the interface transparency and $\phi$ the superconducting phase 
difference between the two leads. For $\phi=0$, it is seen that the bound-state lies right at the gap edge, independent of the the 
interface transparency. As a consistency-check, we have verified that we also obtain this result numerically when using a 
non-selfconsistent step-function profile of the superconducting order parameter. 
Let us now turn to the self-consistent treatment in Fig.~\ref{fig:SNS_total}, where we consider four scenarios for a junction. 
In (a), we have no FVM (i.e.~$\mu_S=\mu_N$) and a short junction, $L_N=8$ sites to be compared with $\xi \sim 25$ sites. The 
superconducting regions are chosen to be large, $L_S=45$ sites, so that they act as superconducting reservoirs. As seen, the 
self-consistent solution for the order parameter only slightly shifts the bound-states inside the gap. This can be understood 
as a consequence of the proximity effect suppression of the order parameter near the interface.
In (b), we increase the length of the normal region to $L_N=24$ sites. In this case the bound-states now reside well within the 
superconducting gap, in contrast to the analytical prediction for ultrashort junctions. Note that the large value of $L_N$ 
ensures that these states do not pertain to some surface-effect, but that they penetrate into the entire N region.
When turning on a FVM ($\mu_S>\mu_N$) in (c) and (d), it is seen that the magnitude of the in-gap bound-states in the N region 
is strongly reduced, due to the reduced normal-state DOS in the N region, although the general energy dependence is seen to be 
similar as in (a) and (b). 

%
\begin{figure}[t!]
\centering
\resizebox{0.49\textwidth}{!}{
\includegraphics{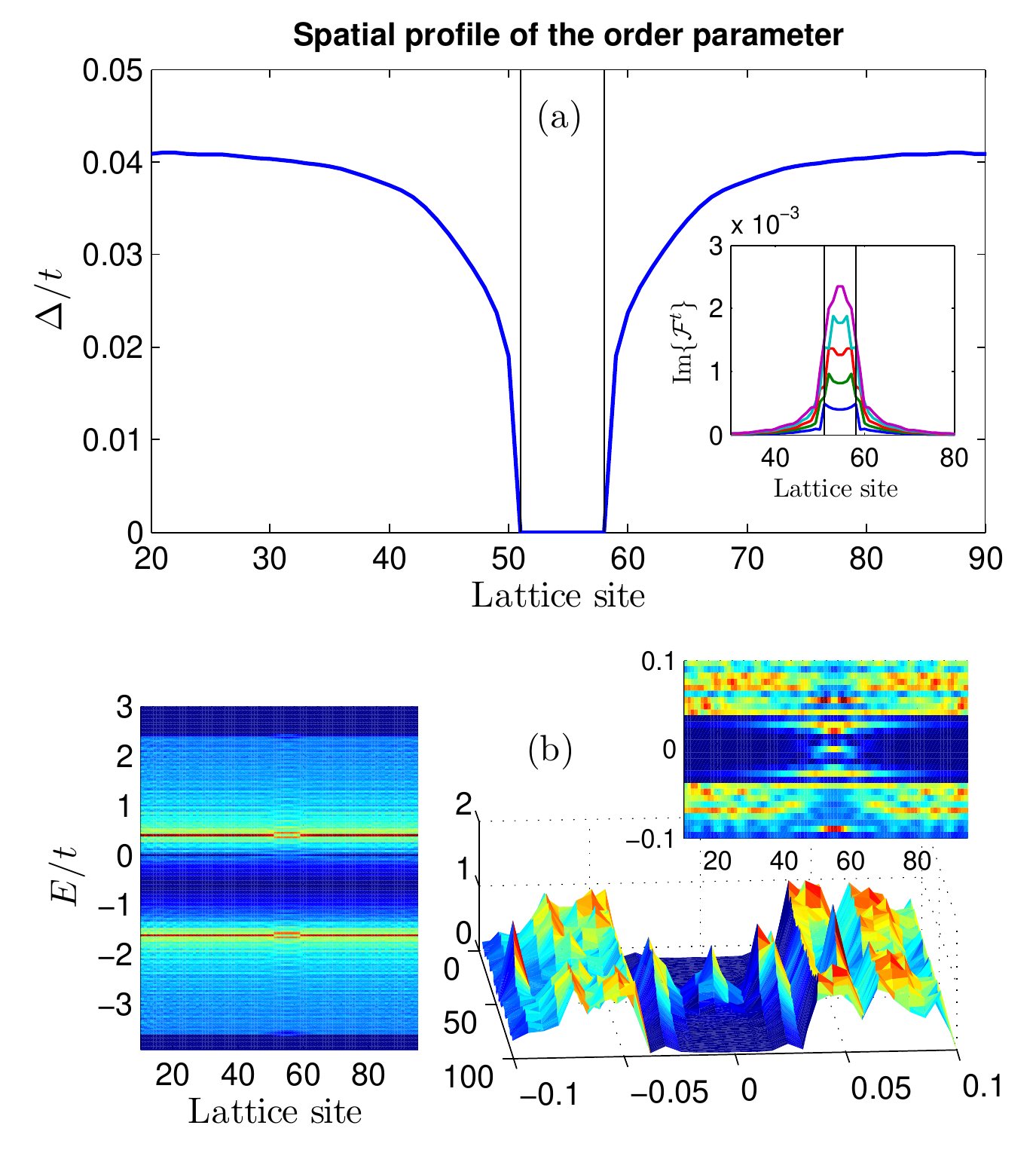}}
\caption{(Color online). Local DOS in an S$\mid$F$\mid$S graphene junction with $L_S=50$ sites, $L_F=8$ sites, $h/t=0.05$, and 
$\mu_S/t=\mu_F/t=0.6$ (no FVM). (a): Spatial profile for the superconducting order parameter. The position of the interfaces are 
marked with black vertical lines. The inset shows the imaginary part of the induced odd-frequency correlations, which from bottom 
to top correspond to times $t\tau=1,2,3,4,5$. (b): Local DOS on a large energy scale (left panel) and near the gapped region 
(right panels). Large values of the DOS are indicated by a bright color, while small values are given by a dark color. As seen,
 odd-frequency correlations give rise to a zero-energy peak in the DOS.}
\label{fig:SFS_short}
\end{figure}
We now turn to the case where the region between the superconducting leads is ferromagnetic via a proximity to a ferromagnetic 
host material. The ground-state of an S$\mid$F$\mid$S junction can occur for a superconducting phase difference $\phi$ of either 
0 or $\pi$, so it becomes necessary to consider the free energy of the junction to correctly identify the ground-state. In 
Fig.~\ref{fig:SFS_short} we consider a short junction with $L_F=8$ sites and no FVM. Our results remain qualitatively the 
same also when including a moderate FVM (e.g.~$\mu_F/\mu_S=0.8$). However, for a sufficiently low chemical potential in the 
F region, the DOS becomes too small to sustain any appreciable in-gap electron density, as was also seen in Fig. \ref{fig:SNS_total}. 
In the experimentally relevant scenario, it is reasonable to expect that $\mu_F$ is doped away from the Dirac point to support 
the presence of ferromagnetism, which is exactly the case considered here.
In Fig. \ref{fig:SFS_short}(a), we show the spatial self-consistent profile of the superconducting order parameter and the 
generation of odd-frequency correlations in the F region (inset). The magnitude of the corresponding anomalous Green's function 
peaks in the middle of the ferromagnetic region, and penetrates a short distance into the S regions. We have here plotted the 
imaginary part of $\mathcal{F}^t$ since it couples directly to the DOS.\cite{yokoyama_prb_07} The experimental manifestation 
of such odd-frequency correlations have previously been discussed in conventional metals, in which case one expects an 
enhancement of the DOS at the Fermi level.\cite{yokoyama_prb_07} In Fig. \ref{fig:SFS_short}(b) we demonstrate that the 
same signature applies for graphene: a zero-energy peak emerges and it is also flanked by additional in-gap bound-states. 
The ground-state for the parameters used in Fig. \ref{fig:SFS_short} was numerically found to be the 0-phase. 

%
\begin{figure}[t!]
\centering
\resizebox{0.49\textwidth}{!}{
\includegraphics{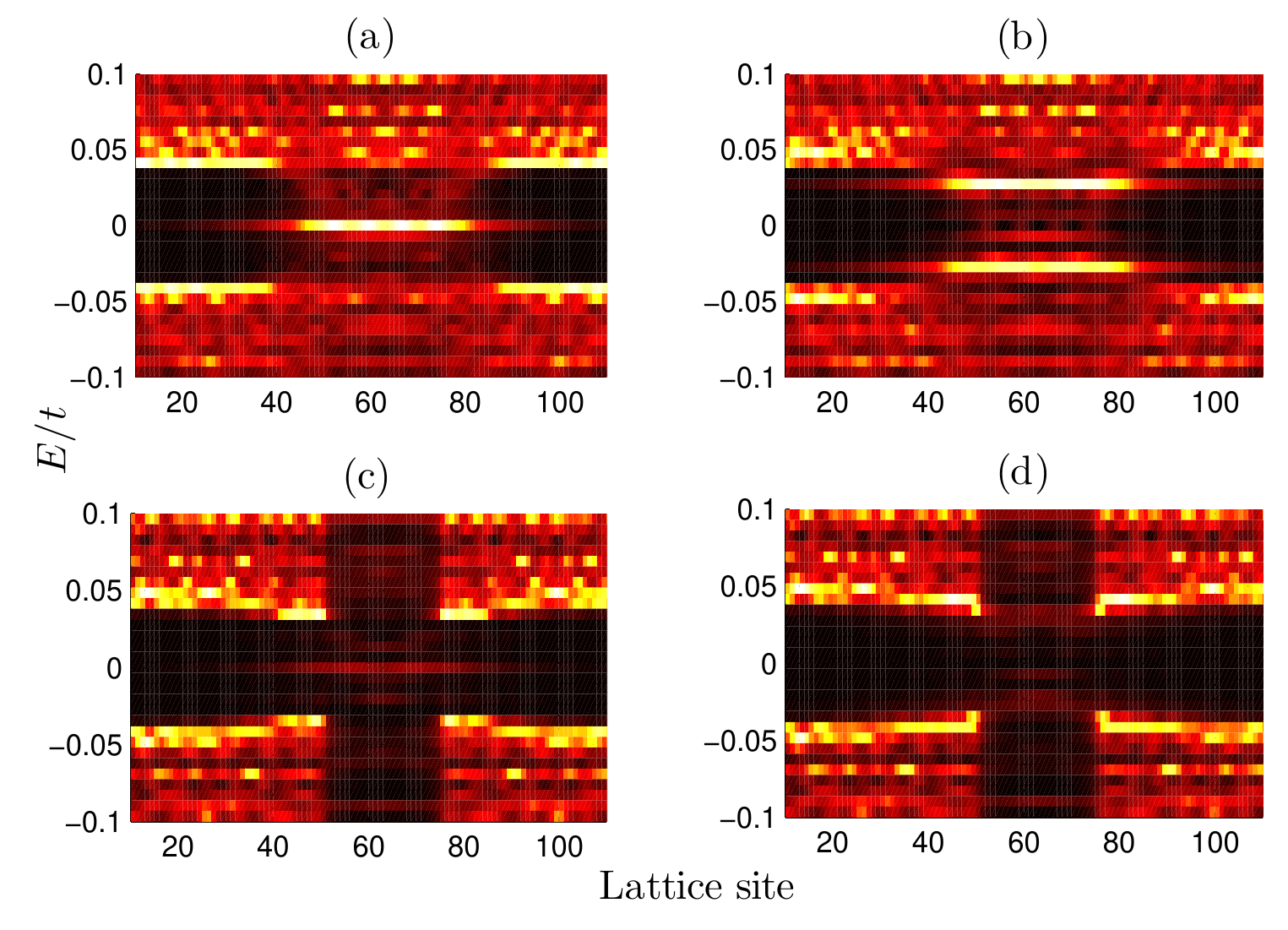}}
\caption{(Color online). Local DOS in an S$\mid$F$\mid$S graphene junction with $L_S=50$ sites, $L_F=24$ sites, $\mu_S/t=0.6$, and 
$h/t=0.06$. (a): 0-phase, $\mu_F/t=0.6$ (no FVM). (b): $\pi$-phase, $\mu_F/t=0.6$ (no FVM). (c): 0-phase, $\mu_F/t=0.2$. 
(d): $\pi$-phase, $\mu_F/t=0.2$. Large values of the DOS are indicated by a bright color, while small values are given by 
a dark color. As seen, the odd-frequency pairing amplitude is most pronounced in the 0-phase, both with and without a FVM.}
\label{fig:SFS_long}
\end{figure}

It is well-known that the superconducting phase-difference $\phi$ can be tuned actively via an external flux or current flowing 
through the system. We investigate in Fig. \ref{fig:SFS_long} how the local DOS in a S$\mid$F$\mid$S junction changes when the 
length of the F region is increased ($L_F=24$ sites) and compare specifically the 0- and $\pi$-phases. As seen in Fig. 
\ref{fig:SFS_long}(a), the odd-frequency pairing correlations are clearly manifested for the 0-phase as a strong zero-energy 
peak when there is no FVM. Going to the $\pi$-phase in (b), it is seen that this peak is suppressed whereas the non-zero 
bound-states inside the gap are instead more pronounced. Upon introducing a strong FVM in (c) and (d), we see again how the 
magnitude of the proximity effect in the F region is severely suppressed, although some signs of the odd-frequency correlations 
are still visible in (c) and (d).

The presence of significant in-gap DOS in the whole F region, even for long junctions ($L_F=24$ sites), rules out the possibility 
of these being caused by surface-states at the S$\mid$F interfaces.
 We have performed numerical calculations for several sets of parameters to investigate the robustness of the odd-frequency peak, and find 
 that it in general competes with the singlet correlations which instead induce a standard mini-gap in the electronic spectrum inside the 
 junction. In spite of this coexistence, our results above demonstrate that the odd-frequency amplitude can be read out from spectroscopic 
 information in a feasible parameter regime. 
 

%
\textit{Summary.} In summary, we have  investigated in a self-consistent manner the proximity effect and its implications for the local 
DOS in both magnetic and non-magnetic graphene Josephson junctions. We have considered several experimentally relevant ranges of doping 
levels and junction lengths. It is found that a considerable triplet proximity effect can be induced in an S$\mid$F$\mid$S graphene 
junction, giving rise to so-called odd-frequency correlations. These are manifested clearly as zero-energy peaks in the DOS, which may 
be probed by STM-measurements. We have also identified the appearance of Andreev-bound states in a S$\mid$N$\mid$S graphene Josephson 
junction. They appear as in-gap resonances in the DOS, in contrast to non-self-consistent results positioning them at the gap edge. Our 
results should be helpful for the interpretation of spectroscopic data and will hopefully serve as a guideline for future experimental 
activity.

\textit{Acknowledgments}. M. Cuoco is thanked for helpful discussions. J.L.
and A.S. were supported by the Norwegian Research Council
Grant No. 167498/V30 (STORFORSK).

\end{document}